\documentclass[twocolumn,showpacs,preprintnumbers,amsmath,amssymb,prl,superscriptaddress]{revtex4-1}

\usepackage{graphicx}% Include figure files
\usepackage{dcolumn}% Align table columns on decimal point
\usepackage{bm}% bold math
\providecommand{\boldsymbol}[1]{\mbox{\boldmath $#1$}}

\begin{document}

\title{The Generalized Quasilinear Approximation: Application to Zonal Jets}

\author{J. B. Marston}
\email{marston@brown.edu}
\affiliation{Department of Physics, Box 1843, Brown University, Providence, RI 02912-1843, USA}

\author{G. P. Chini}
\email{greg.chini@unh.edu}
\affiliation{Department of Mechanical Engineering \& Program in Integrated Applied Mathematics, University of New Hampshire, Durham, NH 03824, USA}

\author{S. M. Tobias}
\email{smt@maths.leeds.ac.uk}
\affiliation{Department of Applied Mathematics, University of Leeds, Leeds, LS29JT, U.K.}

\date{\today}

\begin{abstract}
Quasilinear theory is often utilized to approximate the dynamics of fluids exhibiting significant interactions between mean flows and eddies. In this paper we present a generalization of quasilinear theory to include dynamic mode interactions on the large scales. This generalized quasilinear (GQL) approximation is achieved by separating the state variables into large and small zonal scales via a spectral filter rather than by a decomposition into a formal mean and fluctuations.  Nonlinear interactions involving only small zonal scales are then removed.  The approximation is conservative and allows for scattering of energy between small-scale modes via the large scale (through non-local spectral interactions). We evaluate GQL for the paradigmatic problems of the driving of large-scale jets on a spherical surface and on the $\beta$-plane and show that it is accurate even for a small number of large-scale modes. As this approximation is formally linear in the small zonal scales it allows for the closure of the system and can be utilized in direct statistical simulation schemes that have proved an attractive alternative to direct numerical simulation for many geophysical and astrophysical problems.
\end{abstract}

\pacs{47.27.wg, 47.27.eb, 92.60.Bh, 92.10.A-}

\maketitle

%\section{Introduction}

Even with the advent of peta- and exascale computing, many problems of nonlinear physics are not be amenable to direct numerical simulations (DNS) of the governing partial differential equations (PDEs) in the parameter regimes of physical relevance.  For example, geophysical and astrophysical flows exhibit variability over such a vast range of spatial and temporal scales that DNS of the master PDEs will remain out of reach for the foreseeable future. For this reason a number of complementary approaches to DNS have been investigated that employ approximations of varying complexity.  These approaches generally attempt to achieve some degree of fidelity for the evolution of the large spatial scales, whilst at the same time parameterizing the small-scale interactions in a {\it sub-grid} model; see for instance \citep{FoxKemper:2014ty} and the references therein. Typically, the models are constructed by postulating an {\it ad hoc}, though usually plausible, prescription for the response of the large scales to the small-scale interactions in the form of transport coefficients. 

A more robust approach is to construct {\it self-consistent} equations for the evolution of the low-order statistics of the flow.  Termed Direct Statistical Simulation (DSS), this technique has been shown to be able to reproduce mean flows and two-point correlation functions for model problems describing a wide range of physical processes. In its simplest form DSS employs a quasilinear (QL) approximation to describe the interaction between the large and small scales \cite{farrellioannou2007,Marston:2008tj,sriyoung12}.  The QL approximation has a long history, owing to its utility in the derivation of analytical theories for turbulent interactions and interactions between waves and mean flows; quasilinear equations often also arise naturally as the result of the asymptotic reduction of a master system of PDEs (see  \cite{diamonditohetal2005} and the references therein).  This simplest form of DSS has been utilized successfully to describe the statistics of a range of physical systems including the driving of mean flows in plasmas and on giant planets \cite{tobiasdagonetal2011}, the sustenance of wall-bounded shear-flow turbulence \cite{thomasetal2015}, the growth of a dry atmospheric convective boundary layer \cite{AitChaalal:2015di}, and even the development of the magnetorotational instability in accretion discs \cite{squirebhatt2015}.

Although formally exact in the limit of strong mean flows or for a sufficient separation of timescales \cite{Bouchet2013}, DSS truncated at second-order in the nonlocal equal-time cumulant (denoted S3T \cite{farrellioannou2007} or CE2 \cite{Marston:2008tj}) works less effectively as the system is driven harder, reducing time scale separation with the faster dynamics. This was clearly demonstrated in Ref.~\onlinecite{tobiasmarston2013}, which compared the statistics derived from DNS for the problem of driving $\beta$-plane jets with those calculated by quasilinear DSS (CE2).  CE2 reproduced both the number and strength of jets for large time scale separation but failed as the system was driven further away from equilibrium.

One way to remedy this failure in statistical closures is to employ a higher-order truncation for DSS. This has been pursued  by extending the statistical scheme to include eddy/eddy scattering \cite{Marston:2014vn}. The resulting schemes (termed CE2.5 or CE3$^*$) have a significantly higher computational cost than those that rely on quasilinear approximations, but {\it do} perform better as the system is driven further away from equilibrium \cite{Marston:2014vn}. Perhaps a better approach is to generalize the QL approximation itself so that it remains amenable to analysis {\it and}  provides a foundation for a new statistical method. In this paper we evaluate this generalization of QL, which  we call GQL, for the important problem of the driving of barotropic jets and demonstrate that it can work even in parameter regimes for which QL (and hence CE2/S3T) will fail. These results motivate a new form of DSS that generalizes the second-order cumulant expansion (GCE2) that we shall demonstrate in a subsequent paper.

%\section{Generalized Quasilinear Approximation}

\begin{figure}
\centerline{\includegraphics[width=3.5in]{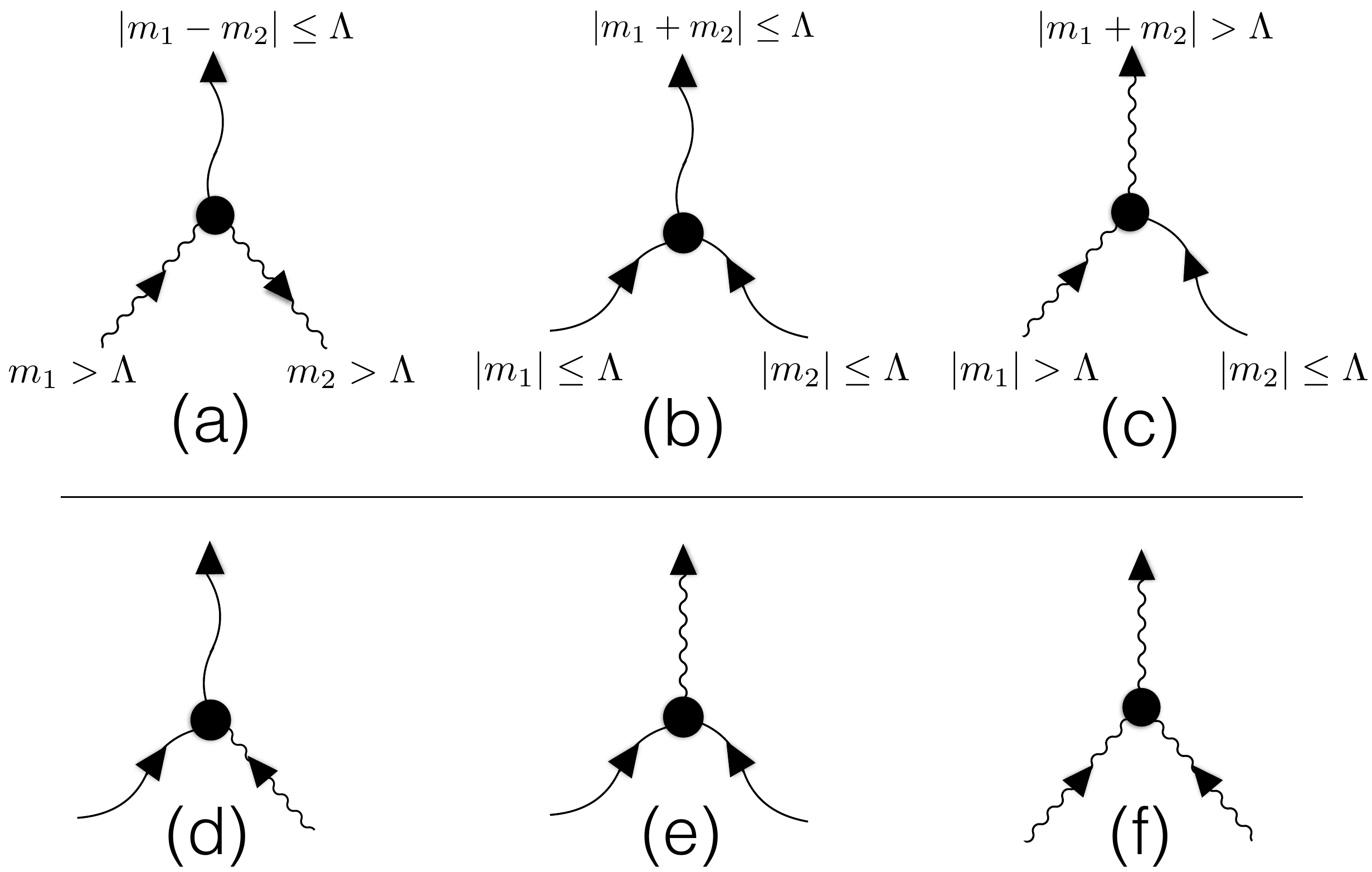}}
\caption{\label{GQLtriads} Triad interactions between modes. Long wavelength fields $\overline{\bf q}$ with zonal wavenumber $|m| \leq \Lambda$ and short wavelength fields $\bf q^\prime$ with $|m| > \Lambda$ are shown. Diagrams (a)--(c) in the top row are retained by the GQL approximation and triad interactions (d)--(f) in the bottom row are omitted.}
\end{figure}

Consider modeling the evolution of a state vector ${\bf q}(\vec{{r}}, t)$ that is specified by a system of master PDEs. For simplicity of exposition, we consider the case where all the nonlinearities in the system are quadratic. This system can be written as 
\begin{equation}
{\bf q}_t = {\cal L}[{\bf q}]+{\cal N}[{\bf q},{\bf q}],
\label{EOM}
\end{equation}
where ${\cal L}$ represents a linear vector differential operator and ${\cal N}$ is the operator that includes the nonlinear (quadratic) interactions such as those in the material derivative.  Specializing to models that are translationally invariant in one direction, denoted the zonal direction, we proceed by generalizing the standard Reynolds decomposition of the state vector into parts, one that oscillates slowly, and one rapidly, in that direction: ${\bf q} = \overline{{\bf q}}+{{\bf q}^\prime}$.  The bandpass filters that we choose are projection operators that obey
$\overline{\overline{\bf q}}=\overline{\bf q}\ \ {\rm and}\ \ \overline{{\bf q}^{\prime}}=0$.
On a rotating sphere, for instance, the bandpass filter separates components that oscillate in the azimuthal ($\phi$) direction at zonal wavenumbers  $|m|$ less than or equal to $\Lambda$ from those that oscillate with $|m|>\Lambda$:
%less than or equal to $\Lambda$ from those greater than $\Lambda$:
%On a rotating sphere, for instance, the bandpass filter separates components that oscillate in the azimuthal ($\phi$) direction at zonal wavenumbers less than or equal to $\Lambda$ from those greater than $\Lambda$:
\begin{eqnarray}
\overline{\bf q}(\theta, \phi)  &=& \sum_{|m| \leq \Lambda} e^{i m \phi}~ {\bf q}_m(\theta),
\nonumber \\
{\bf q}^\prime(\theta, \phi)  &=& \sum_{|m| > \Lambda} e^{i m \phi}~ {\bf q}_m(\theta). 
\end{eqnarray} 
The GQL approximation is then obtained by neglecting interactions in the evolution equations as described by Figure~1; hence
\begin{eqnarray}
{\overline{\bf q}}_t &\equiv& {\cal L}[\overline{\bf q}] + \overline{{\cal N}}[{{\bf q}^{\prime},~{\bf q}^{\prime}}] + \overline{{\cal N}}[\overline{\bf q},~ \overline{\bf q}]
\label{GQLEOMa}\\
{{\bf q}^{\prime}}_t &\equiv& {\cal L}[{\bf q}^{\prime}] + {\cal N}^\prime[\overline{\bf q},~ {{\bf q}^{\prime}}] + {\cal N}^\prime[{{\bf q}^{\prime}},~ \overline{\bf q}]\ .
\label{GQLEOMb}
\end{eqnarray}
The two nonlinear terms that appear in Eq. \ref{GQLEOMa} correspond to diagrams (a) and (b) in Fig. \ref{GQLtriads} and the sum of the two nonlinear terms in Eq. \ref{GQLEOMb} is interaction (c).  
Terms $\overline{\cal N}[\overline{\bf q},~{\bf q}^\prime] + \overline{{\cal N}}[{\bf q}^\prime,~ \overline{\bf q}]$ [diagram (d)], ${\cal N}^\prime[\overline{\bf q},~ \overline{\bf q}]$ [(e)], and ${\cal N}^\prime[{\bf q}^\prime,~ {{\bf q}^{\prime}}]$ [(f)] are discarded. 
The QL approximation is recovered in the limit $\Lambda = 0$, for which $\overline{\bf q}$ is simply the zonal mean.  In the opposite limit $\Lambda \rightarrow \infty$, ${\bf q}^\prime = 0$ and the exact and fully nonlinear (NL) dynamics of Eq. \ref{EOM} are recovered.  Thus GQL interpolates between QL and the exact dynamics and provides a systematic way to improve the QL approximation.  

A particularly important feature of the new GQL system is that the triad interactions that are retained respect the linear and quadratic conservation laws of the original model such as conservation of angular momentum, energy and enstrophy. Two advantages of GQL theory over QL theory should be emphasized.  First, the low modes are allowed to undergo fully nonlinear interactions and thus constitute the resolved model.  Satellite modes or zonons therefore can be captured \cite{Bakas:2013bo}.  The second advantage is that small-scale eddies can exchange energy through their interaction with the large scales; that is, in a GQL system energy can be redistributed among smaller scales via scattering off the large-scale flows, a non-local spectral transfer.  The high-modes thus function as a novel, deterministic sub-grid model. Moreover, GQL is able to predict the responses of the {\it distributions} for quadratic fluxes (here, Reynolds stresses) in a manner that simply is not possible for QL, since this energy redistribution is {\it not} captured by a QL (or equivalently, CE2) description.  Given that these responses are crucial to the successful description of the generation of large-scale fields, this is a key step.

%\section{Testing GQL: Driving of Zonal Jets on the Sphere and $\beta$-plane}

We examine the effectiveness of the GQL approximation as applied to barotropic dynamics on a spherical surface ($0 \le \phi <  2 \pi$, $0 \le \theta \le  \pi$) and on a local Cartesian $\beta$-plane ($0 \le x <  2 \pi$, $0 \le y <  2 \pi$). The barotropic vorticity equation for $\zeta \equiv {\bf e}_1
\cdot (\boldsymbol{\nabla} \times \boldsymbol{u})$, where ${\bf e}_1 = \hat{\bf r}$ on the spherical surface and  ${\bf e}_1 = \hat{\bf z}$ on the $\beta$-plane, is given by 
\begin{equation}
\partial_t \zeta + J(\psi,f+\zeta)  = -\kappa \zeta+{\cal D}+\eta(t). \label{z_eqn1}
\end{equation}
Here $J(~,~)$ is the Jacobian and we have utilized the streamfunction representation $\boldsymbol{u}= \boldsymbol{\nabla} \times (\psi {\bf e}_1)$.
The motion is damped by Rayleigh friction --$\kappa\zeta$ and also a dissipation ${\cal D}$ that removes small-scale structures. 
On the spherical surface ${\cal D}$ takes the form of a hyperdiffusion, whilst on the $\beta$-plane viscous dissipation $\nu \nabla^2 \zeta$ is used. Finally, rotation is incorporated through the Coriolis parameter $f$, where $f=2 \Omega \cos \theta$ for the spherical system and $f=f_0+\beta y$ for the local $\beta$-plane model. The forcing $\eta(t)$ is chosen to be stochastic and narrow band in spectral space, with a short renewal time \cite{Marston:2014vn}.

Much is known about the dynamics of this system (often referred to as the Charney--Hasagawa--Mima system) owing to its importance as a paradigm problem for the formation of jets in stably stratified fluid environments such as the Earth's atmosphere and oceans, the outer layers of gas giants, stellar interiors and exoplanets (see for instance \cite{rhin75,vallis06, galetal2010} and the references therein). This system also is used as a simple model the formation of zonal flows in tokamaks (see \cite{diamonditohetal2005}). Briefly, energy injected at small to moderate scales is transferred to larger scales via nonlinear interactions.  In the fluid system the energy transfer occurs via the nonlinear interaction of Rossby waves, whilst in the plasma system drift waves facilitate the transfer. Owing to the underlying anisotropy of the system, the energy at large scales is preferentially transferred into systematic zonal flows (here a zonal flow is in the $\phi$- or $x$-direction). Two competing mechanisms have been proposed for the energy transport.  The first involves the scale-by-scale transfer of energy known as the inverse cascade \cite{valmal93}, in which small-scale/small-scale interactions lead to the deposition of  energy  in other small scales.  Via successive interactions, this energy is  able to make its way to the largest scale; the process therefore involves eddy/eddy scattering. The other mechanism is non-local in wavenumber space and relies upon the direct transfer of energy to the largest scales.  This forcing via Reynolds stress terms only involves eddies interacting with eddies directly to produce mean flows. Of course in any real fluid system both of these mechanisms are operative, with their relative importance often characterized by the Kubo number or its analogues \cite{diamonditohetal2005}.

%\subsection{Spherical Surface}

On the sphere, pure spectral DNS with truncation in wavenumbers $0 \leq \ell \leq L$ and $|m| \leq \min\{\ell, M\}$ is performed.  We choose spectral cutoffs $L = 30$ and $M =20$  and work on the unit sphere and in units of time (days) and thus $\Omega = 2 \pi$.  
To remove enstrophy cascading to small scales, hyperviscosity $\nu_3 (\nabla^2+2) \nabla^4 \zeta$ is included in the linear operator in Eq.~\ref{EOM}.   
We choose parameters for the jet as in Ref.~\onlinecite{Marston:2014vn}.  The fluid motion is driven by stochastic forcing $\eta$ and damped by friction with $\kappa = 0.02$.   Only modes with $8 \leq \ell \leq 12$ and $8 \leq |m| \leq \ell$ are stochastically forced (with $|m| \leq \ell$).  This has the effect of confining the stochastic forcing to lower latitudes.

\begin{figure}
\centerline{\includegraphics[width=3.4in]{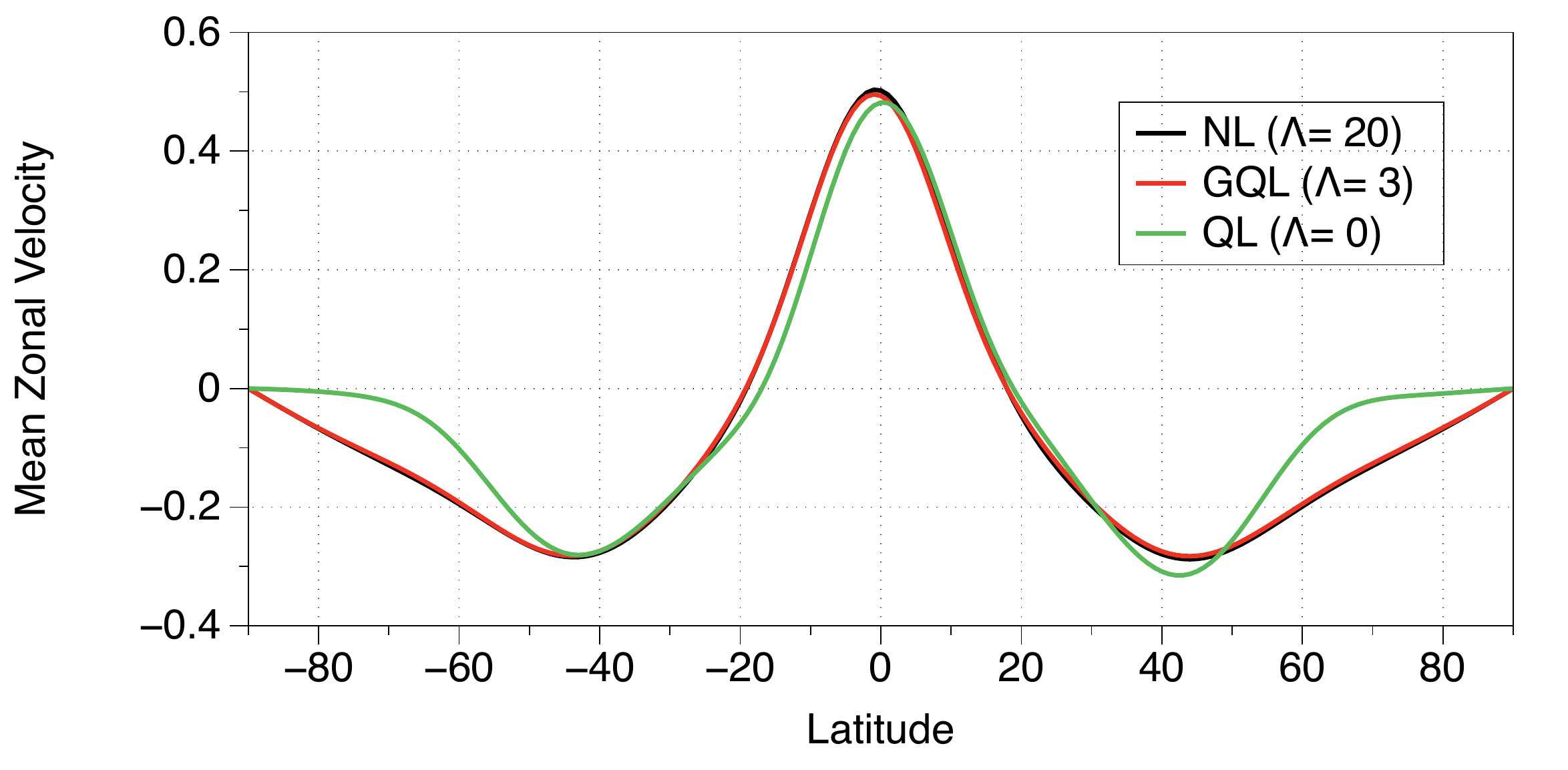}}
\caption{\label{figure2} Time- and zonal-averaged zonal velocity as a function of latitude.  Time averaging over 5,000 days commences after a spin-up of 500 days.  NL is the fully nonlinear (NL) simulation.  In the QL limit of $\Lambda = 0$ there is no mechanism to transfer angular momentum from low latitudes, where it is forced, to high latitudes.  GQL with $\Lambda = 3$ corrects this defect.}
\end{figure}  

\begin{figure}
\centerline{\includegraphics[width=3.4in]{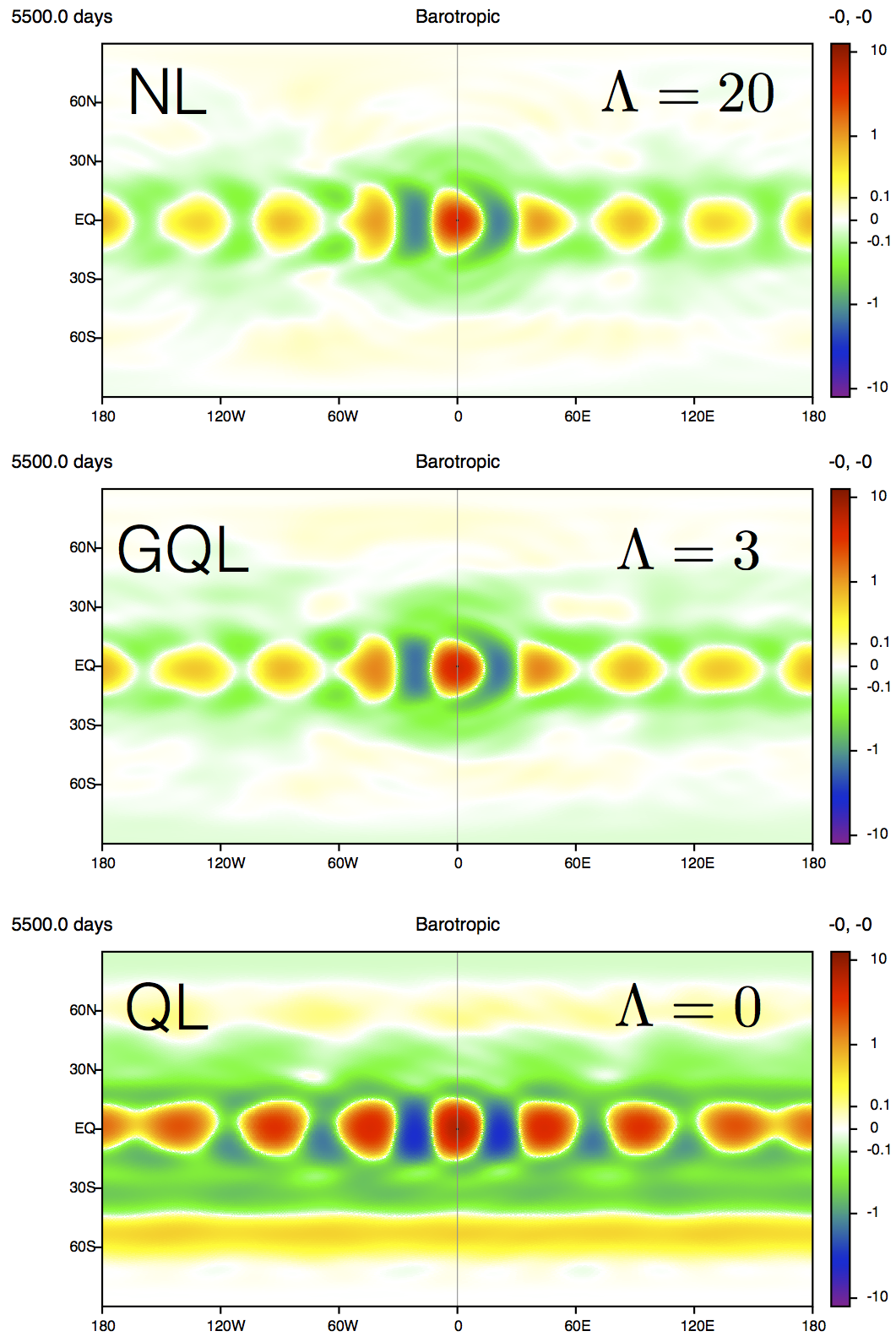}}
\caption{\label{figure3} Second cumulant (two-point correlation function of the vorticity).  One point is centered along the prime meridian and at latitude $0^\circ$.  The non-local nature of the correlations (or `teleconnections') is evident.  As in Figure \ref{figure2}.}
\end{figure}  

Figure \ref{figure2} shows that, in contrast with QL, GQL is in fact able to scatter angular momentum to high latitudes, reproducing the zonal flow there as well as at low latitudes.  Figure \ref{figure3} similarly confirms that GQL is able to reproduce the two-point correlation function of the vorticity, again in contrast to QL ($\Lambda = 0$), which shows too strong and too coherent waves owing to the absence of eddy-eddy scattering.

%\subsection{$\beta$-plane}

Next we present an evaluation of GQL on the $\beta$-plane for various choices of the cutoff wavenumber $\Lambda$ by comparing it with DNS (for which $\Lambda \rightarrow \infty$).
We start by performing a full DNS using a pseudo-spectral scheme optimized for use
on massively parallel machines \cite{tobcatt08} at a spectral resolution of $2048^2$.
As in \cite{tobiasmarston2013} the stochastic forcing is chosen to be random but here concentrated in a spectral band of
wavenumbers  $11 \le |k_x|, |k_y| \le 14$. The system is evolved from a state of rest until a solution with 4 strong jets is reached. This state is moderately far from equilibrium as measured by the so-called Zonostrophy Index $R_\beta = 2.6$ \cite{galsukdik08}. 
It is the subsequent evolution of this state under the various degrees of approximation (different choices of $\Lambda$) that we shall investigate here.

Figure~4(a) shows the evolution of the zonal jets as a function of latitude ($y$) and time ($t$) in a 
Hovm\"oller diagram for the mean vorticity $\overline{\zeta}(y,t)$. At the start of the interval 4 jets are clearly visible, but this configuration is unstable and just after half-way through the evolution the two jets nearest the top of the computational domain merge to form one jet. This complicated merging process occurs relatively quickly and the three jet structure remains stable until the end of the computation. Figure~4(b-f) show the comparable evolution in space and time of the mean flow for GQL at various levels of truncation. Figure~4(b,c) illustrate the dynamics for $\Lambda=50$ and $\Lambda=10$, respectively; for these cases, the large-scale modes are driven directly.
It is clear that these truncations {\it are} capable of reproducing the jet merging process and the continued evolution of the three-jet solutions. Figure~4(d-f) show Hovm\"oller diagrams for the cases $\Lambda=5$, $3$ and $0$.  For these cases the forcing injects energy directly into small-scale modes.  The large scales can receive energy only from small-scale/small-scale interactions; this energy then can be redistributed among the large-scale modes through their dynamic self-interaction (except in the case where $\Lambda=0$). Remarkably, the GQL approximation run at $\Lambda=5$ and even $\Lambda=3$ is able to accurately reproduce the large-scale dynamics of the fully NL DNS. It is only for the quasilinear system ($\Lambda=0$) that the qualitative dynamics are not replicated; for this case only two jets remain at the end of the run, and the transitions to reach this state are significantly different.  The efficacy of the GQL over the QL approximation also may be demonstrated by examining the two-dimensional spectra when averaged over the second half of the evolution. It is clear from Figure~5 that GQL is able to redistribute power over a wide range of zonal wavenumbers.

\begin{figure}
\centerline{\includegraphics[width=3.5in]{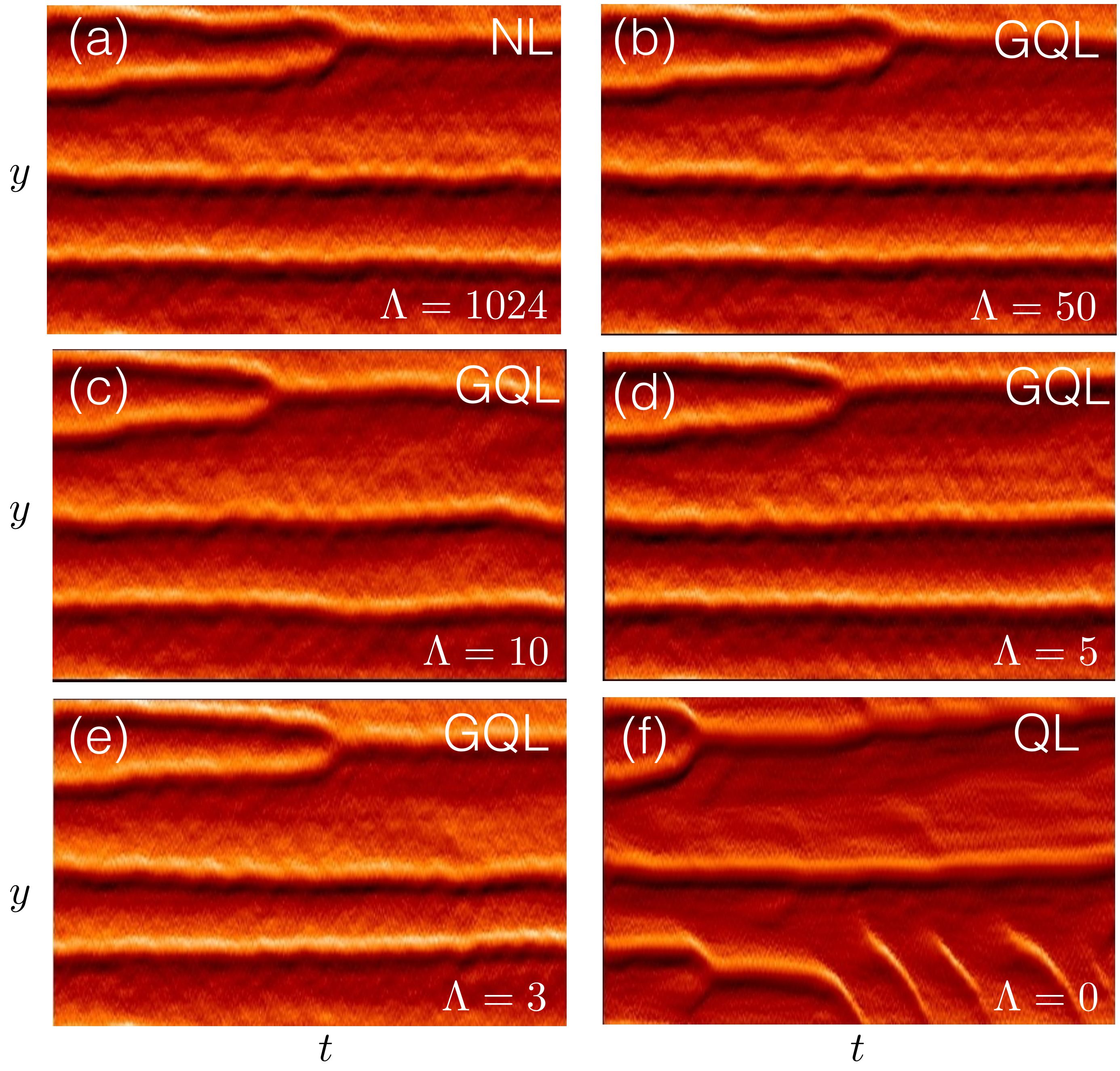}}
\caption{\label{figure4} Color-coded Hovm\"oller diagram (space-time plots with time $t$ along the horizontal axis and $0  \le y \le 2 \pi$ along the vertical axis) for the mean vorticity $\overline{\zeta}(y,t)$ for (a) NL DNS ($\Lambda = 1024$), (b) $\Lambda = 50$, (c) $\Lambda = 10$, (d) $\Lambda = 5$, (e) $\Lambda = 3$, (f) $\Lambda = 0$ (equivalent to QL).}
\end{figure}  

\begin{figure}
\centerline{\includegraphics[width=3.5in]{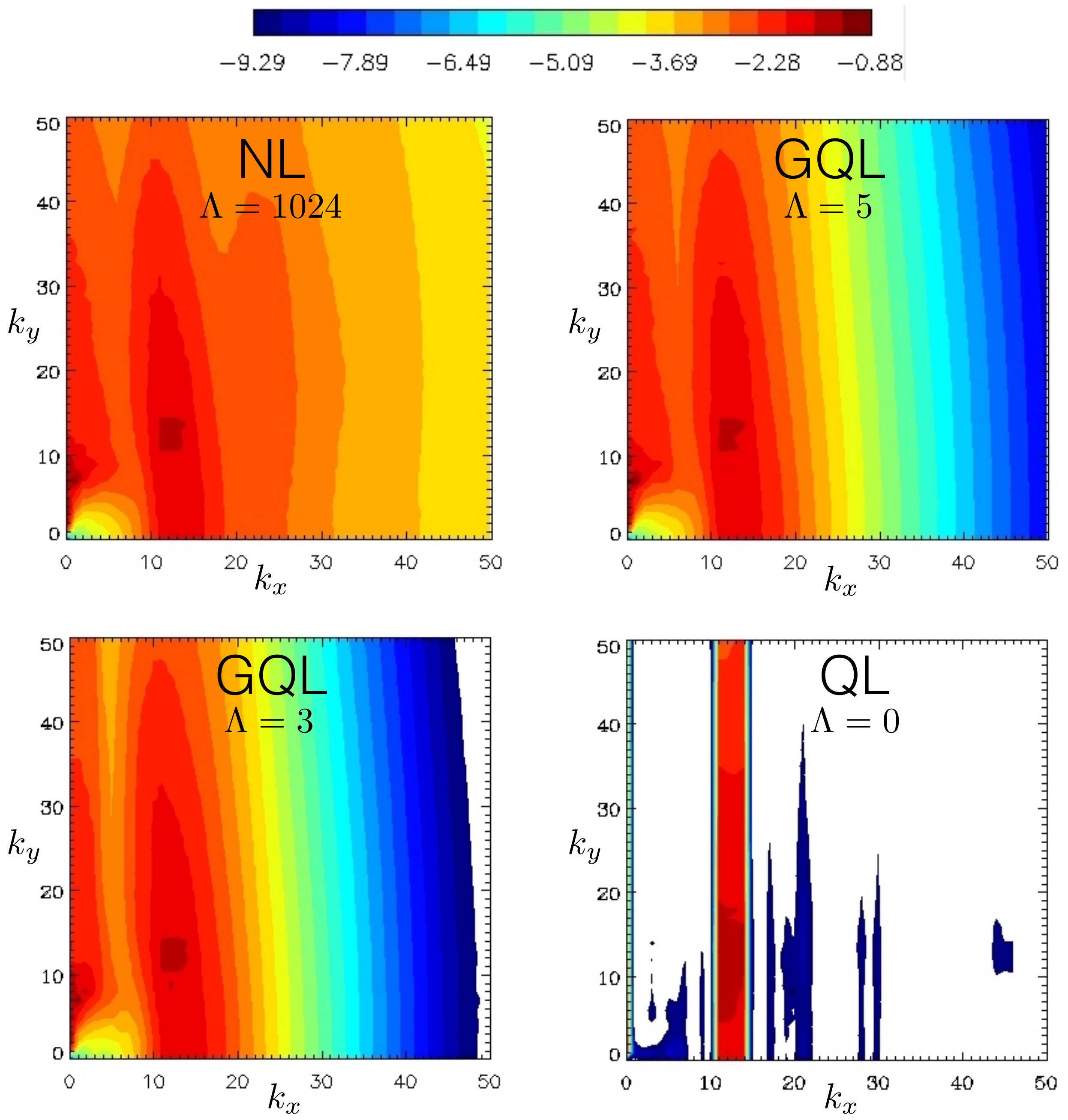}}
\caption{\label{figure5} Time-averaged vorticity power spectra on a $\log_{10}$ scale for fully NL DNS ($\Lambda = 1024$), GQL with $\Lambda = 5$ and $\Lambda = 3$, and QL ($\Lambda = 0$).  Like NL, GQL redistributes power throughout spectral space.}
\end{figure}  

%\section{Discussion}

A formal justification for the GQL approximation is provided by a multiple-scales asymptotic reduction of the PDEs governing a particular class of anisotropic incompressible flows. Briefly, in the context of the barotropic $\beta$-plane vorticity equation (Eq.~\ref{z_eqn1}), {\it two} zonal coordinates and time scales are introduced:  $\chi \equiv x$ and $X \equiv \epsilon x$, and $\tau \equiv t$ and $T \equiv \epsilon t$, where $\epsilon$ is a formal scale separation parameter that can be related to the ratio of dissipation to forcing. 
A {\it fast-averaging} operation over ($\chi$,$\tau$) is introduced such that each dependent variable can be decomposed, in physical space, into a coarse-grained, {\it slowly-varying} mean field (again denoted with an overbar) and a fluctuation (denoted with a prime):   $\zeta(\chi,X,y,\tau,T;\epsilon)=\overline{\zeta}(X,y,T;\epsilon)+\zeta'(\chi,X,y,\tau,T;\epsilon)$.  Eq.~(\ref{z_eqn1}) with multiscale derivatives is parsed into equations for the mean and fluctuation fields, and the vorticity $\zeta$ and streamfunction $\psi$ are expanded in asymptotic series in (fractional) powers of $\epsilon$ as:
$\zeta\sim\overline{\zeta}_0+\sqrt{\epsilon}\left(\overline{\zeta}_1+\zeta_1^\prime\right)+\ldots,$ and similarly for $\psi$.
The form of these expansions is dictated by the requirements that:  (i)~the large-scale flow be incompressible; (ii)~the (dimensionless) jets $\overline{u}\equiv\partial_y\overline{\psi}$ have $\mathit{O}(1)$ magnitude; and, crucially, (iii)~the meridional Reynolds-stress divergence arising from the fluctuation fields feeds back on the slowly-varying mean.  Given these considerations the fluctuation equations are readily deduced to be {\it quasilinear} about the $\mathit{O}(1)$ mean flow.  The equations for the evolution of the slowly-varying mean fields are obtained via a secularity condition, which yields the system 
\begin{eqnarray}
\partial_T\overline{\zeta}  +J(\overline{\psi},\overline{\zeta}) + \overline{\beta}\partial_X\overline{\psi} &=& - \partial_y\left(\overline{\zeta^\prime\partial_\chi\psi^\prime}\right) - \overline{\kappa}\overline{\zeta} +\overline{\mathcal{D}},\label{zMEANeqn}\\
\partial_y^2\overline{\psi}&=&\overline{\zeta},
\end{eqnarray}
where $\overline{\beta}$, $\overline{\kappa}$ and $\overline{\mathcal{D}}$ are the $\mathit{O}(1)$ rescaled (dimensionless) beta coefficient, Rayleigh friction coefficient and dissipation, respectively, and the subscripts on the leading-order fields have been omitted.  The corresponding leading-order fluctuation system (again dropping subscripts) is given by
\begin{eqnarray}
\partial_\tau{\zeta}^\prime-\partial_y\overline{\psi}\partial_\chi\zeta^\prime+\partial_\chi\psi^\prime\partial_y\overline{\zeta}+\overline{\beta}\partial_\chi\psi^\prime
&=&\eta^\prime(\tau),\\
\left(\partial^2_\chi+\partial_y^2\right)\psi^\prime&=&\zeta^\prime\label{psiFLUCTeqn},
\end{eqnarray}
where $\eta^\prime(\tau)$ is the fluctuation forcing.  Upon reverting to a single set of temporal scales, the GQL formulation can be interpreted as a spectral space implementation of the multiscale reduced PDE system (\ref{zMEANeqn})--(\ref{psiFLUCTeqn}).  In the appropriate asymptotic limit, this alternative reduction not only provides a formal mathematical justification for the GQL formulation but also is suggestive of other multiscale algorithms for simulating the reduced dynamics \citep{malecha:2014}.

An extension of quasilinear theory has been introduced for the reduced description of out-of-equilibrium anisotropic fluid systems.  Instead of incorporating interactions corresponding to the retention of higher-order cumulants, the generalized quasilinear (GQL) theory proposed here is developed by reinterpreting the underlying linearization.  Specifically, in GQL the dynamics is linearized about a coarse-grained rather than strictly mean field that undergoes fully nonlinear interactions.  One crucial consequence of this extension is that energy can be redistributed among small-scale modes via scattering off the large-scale flow, in stark contrast to QL dynamics.  The potential utility of GQL theory has been demonstrated for the canonical problem of the driving of zonal jets in barotropic turbulence; for the parameter regime investigated, the accuracy of the QL method is shown to be significantly improved by retaining even just three coarse modes.  Further advantages of the method include its relative ease of implementation and guaranteed preservation of conservation laws of the master equations.  The GQL formulation thus provides a novel  and seamlessly integrated closure for sub-grid dynamics that should have broad applicability to anisotropic turbulent flows arising in nature and technology.

\begin{acknowledgments}
We wish to acknowledge useful discussions with  Freddy Bouchet, Keith Julien and Baylor Fox-Kemper.
This work was supported in part by NSF under grant Nos. DMR-0605619 and CCF-1048701 (JBM) and CBET-1437851 (GPC).   JBM and SMT thank the Isaac Newton Institute in Cambridge UK for its hospitality during the program ``Mathematics for the Fluid Earth'' where some of this work was carried out.  GPC and JBM thank the WHOI 2015 Summer GFD program for its support.  
%We wish to acknowledge useful discussions with  Freddy Bouchet and Baylor Fox-Kemper.
%This work was supported in part by NSF under grant Nos. DMR-0605619 and CCF-1048701 (JBM).   JBM and SMT thank the Isaac Newton Institute in Cambridge UK 
%for its hospitality during the program ``Mathematics for the Fluid Earth'' where some of this work was carried out.  GPC and JBM thank the WHOI 2015 Summer GFD program for its support.  
\end{acknowledgments}

\bibliography{gql}

\end{document}